\documentclass[a4paper,11pt]{article}

\usepackage{pos}
\usepackage{lineno}
\usepackage{booktabs}
\usepackage{multirow}

\usepackage{lipsum} 

\title{Search for Dark Matter Decay in Nearby Galaxy Clusters and Galaxies with IceCube}

\ShortTitle{Search for Dark Matter Decay in Nearby Galaxy Clusters and Galaxies with IceCube}

\author{The IceCube Collaboration \\{\normalsize \normalfont(a complete list of authors can be found at the end of the proceedings)}\\}

\emailAdd{minjin.jeong@g.skku.edu}
\emailAdd{rott@physics.utah.edu}

\abstract{

Dark matter could decay into Standard Model particles producing neutrinos directly or indirectly. The resulting flux of neutrinos from these decays could be detectable at neutrino telescopes and would be associated with massive celestial objects where dark matter is expected to be accumulated. Recent observations of high-energy astrophysical neutrinos at IceCube might hint at a signal produced by the decay of TeV to PeV scale dark matter. This analysis searches for neutrinos from decaying dark matter in nearby galaxy clusters and galaxies. We focus on dark matter masses from 10 TeV to 1 EeV and four decay channels: $\nu\bar{\nu}$, $\tau^{+}\tau^{-}$, $W^{+}W^{-}$, $b\bar{b}$. Three galaxy clusters, seven dwarf galaxies, and the Andromeda galaxy are chosen as targets and stacked within the same source class. A well-established IceCube data sample is used, which contains 11 years of upward-going track-like events. In this contribution, we present preliminary results of the analysis.

\vspace{4mm}
{\bfseries Corresponding authors:}
Minjin Jeong$^{1*}$, Carsten Rott$^{1,2}$\\
{$^{1}$ \itshape Department of Physics, Sungkyunkwan University, South Korea}\\
{$^{2}$ \itshape Department of Physics \& Astronomy, University of Utah, USA}\\
\\[4mm]
$^*$ Presenter

\ConferenceLogo{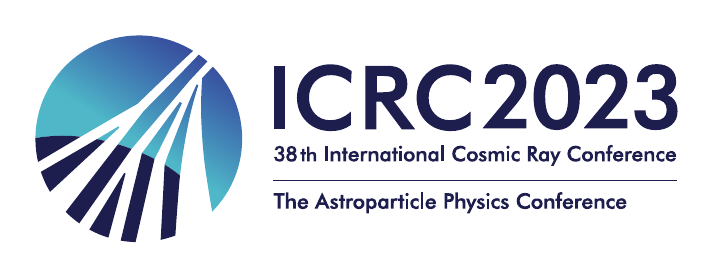}

\FullConference{The 38th International Cosmic Ray Conference (ICRC2023)\\ 26 July -- 3 August, 2023\\ Nagoya, Japan}
}

\begin{document}
\maketitle
\section{Introduction}\label{sec1}

Although the existence of dark matter is well known from its gravitational effects, the nature of dark matter still remains mysterious. Weakly Interactive Massive Particles (WIMPs) arising from extensions to the Standard Model of particle physics have been popular dark matter candidates and are predicted to have masses from a few GeV to a few tens of TeV. Thus, classical dark matter searches focus on the GeV-TeV scale. However, despite decades of efforts, WIMPs have eluded detection. Recently, interests in alternative dark matter models have increased, some of which are motivated by modern neutrino and gamma-ray experiments that can probe PeV-scale phenomena. In particular, it has been speculated that the high-energy astrophysical neutrinos observed at IceCube might hint at a signal from the decay of TeV-PeV dark matter~\cite{Chianese:2017nwe,Esmaili:2013gha, Esmaili:2014rma, Rott:2014kfa}. Recent IceCube analyses derived one of the best bounds on the dark matter lifetime on the TeV-PeV scale, by looking for neutrinos from Galactic and cosmological dark matter decay~\cite{IceCube:HESE7.5DM, IceCube:EPJC2018DM}. We expand previous searches for dark matter with IceCube by searching for signals from decaying dark matter in nearby galaxy clusters and galaxies. In this contribution, we present the analysis methods and preliminary results. 

\section{Neutrinos from Dark Matter Decay}
When dark matter decays in an astrophysical object with a redshift of approximately zero, the expected neutrino flux at Earth's surface reads 
\begin{equation}
\frac{d\Phi_{\nu}}{dE_{\nu}} 
    = \frac{1}{4\pi{m}_{\chi}{\tau}_{\chi}} \frac{dN_{\nu}}{dE_{\nu}} \int_{\Delta \Omega} d\Omega \int_{l.o.s}\rho_{\chi}dl,
\label{eq:signal_flux}
\end{equation}
where $m_{\chi}$ and $\tau_{\chi}$ are the dark matter mass and lifetime, respectively, and $dN_{\nu}/dE_{\nu}$ is the neutrino spectrum expected at Earth's surface. The dark matter mass density distribution ($\rho_{\chi}$) is often assumed to be spherically symmetric and is integrated along the line-of-sight (l.o.s) and over the solid angle ($\Delta\Omega$) corresponding to the region of interest (ROI). Since neutrino events and anti-neutrino events are not distinguishable at IceCube on event-by-event basis, the sum of neutrino and anti-neutrino fluxes is used to estimate the expected signal. 

\begin{figure}[t]
    \centering
    \includegraphics[width=0.5\linewidth]{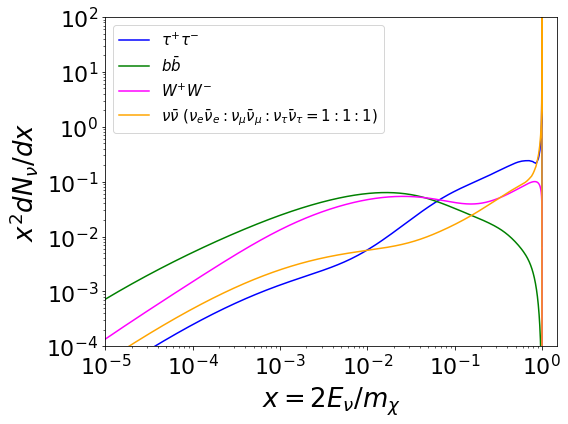}
    \caption{\textbf{Expected neutrino spectra for four different dark matter decay channels.} Each line represents the neutrino spectrum expected from the decay of a dark matter particle of mass 25~PeV. The spectrum is summed over neutrinos and anti-neutrinos, and averaged between the three neutrino flavors. The spectra are presented in terms of an energy fraction ($x=2E_{\nu}/m_{\chi}$) that can be as large as 1.0. The spectrum for the $\nu\bar{\nu}$ channel is calculated assuming flavor democracy. For this channel, a delta function is expected at $x=1$ due to direct decays of dark matter into neutrinos. The low energy tail is expected from the neutrino regeneration via the emission of W, Z bosons from energetic neutrinos.}
\end{figure}

The neutrino spectrum depends on the dark matter mass and decay channel. In this analysis, we consider dark matter masses ranging from 10 TeV to 1 EeV and choose 26 mass values evenly spaced on a logarithmic scale to sufficiently cover the entire mass range. To make the analysis independent of particular dark matter decay models, we assume that dark matter decays into a pair of Standard Model particles with 100\% branching ratio and consider four different decay channels: $\nu\bar{\nu}$, $\tau^{+}\tau^{-}$, $W^{+}W^{-}$, $b\bar{b}$. For a given dark matter mass, the $b\bar{b}$ channel would yield a soft spectrum, i.e., a large fraction of the spectrum is focused on low energies, as neutrinos should be produced via hadronization. On the other hand, the $\tau^{+}\tau^{-}$ channel would yield a hard spectrum, as the tau leptons can directly decay to neutrinos and charged leptons. We use the $\chi aro \nu$ package~\cite{Charon} to calculate the expected neutrino spectra at the production site, and account for neutrino oscillations by averaging the neutrino flavor ratio to 1:1:1. Figure~1 shows the spectra calculated for 25 PeV dark matter mass for the four decay channels. In the horizontal and vertical axes, the energy fraction ($x$) is defined by $x = 2E_{\nu} / m_{\chi}$. In general, the shape of a neutrino spectrum from dark matter decay does not change drastically with the dark matter mass. The total number of neutrinos per dark matter decay can be obtained by converting the $dN_{\nu}/dx$ to $dN_{\nu}/dE_{\nu}$ and integrating it over the energy. For the $\nu\bar{\nu}$ channel, we assume that dark matter particles can decay into one of the  $\nu_{e}\bar{\nu}_{e}$, $\nu_{\mu}\bar{\nu}_{\mu}$, or $\nu_{\tau}\bar{\nu}_{\tau}$ final states each with a one in three chance. This scenario is referred to as flavor democracy. The spectrum from this channel includes a delta function expected from direct decays of dark matter into neutrinos. A continuum for $x < 1$ is also expected due to neutrino regeneration via the emission of W and Z bosons from energetic neutrinos~\cite{Bauer:HDMSpectra_2020jay}.

The double integral on the right hand side of Eq.~(\ref{eq:signal_flux}) is referred to as the D-factor and depends on the dark matter mass distribution in the target. The D-factor is considered an important criterion when selecting targets, as the signal neutrino flux is proportional to it. Nearby galaxy clusters and galaxies whose dark matter halo models are available are considered candidate targets, and among those we select the sources with relatively large D-factor values. Furthermore, we focus on the northern sky from which the atmospheric muons are absorbed by the Earth and cannot reach the IceCube detector located at the geographic South Pole.  

Reference~\cite{SanchezConde:2011ap} presents dark matter halo models, based on the Navarro-Frenk-White (NFW) profile~\cite{NFW:1996}, for nearby galaxy clusters. Among the galaxy clusters discussed in the paper, we select those in the northern sky: the Virgo, Coma, and the Perseus galaxy clusters. The Andromeda galaxy (M31) is also chosen as a target, due to its proximity and large dark matter content. We take the NFW dark matter halo model for this galaxy presented in Ref.~\cite{Tamm:2012hw}. Reference~\cite{Geringer-Sameth:2014} describes  dark matter halo models for twenty dwarf spheroidal galaxies around the Milky Way, using the Zhao profile~\cite{Zhao:1995cp}. Among those, we select seven dwarf galaxies which are located in the northern sky and have relatively large D-factor values. In this work, we stack the galaxy clusters and dwarf galaxies within the same source class in order to maximize the analysis sensitivity and mitigate the impact of their halo model uncertainties, and look for neutrinos from dark matter decay in the three source groups separately. Table~\ref{table:sources} shows properties of the selected targets. In the table, $\theta_{ROI}$ represents the angular distance from the source center up to which the D-factor ($D_{ROI}$) is calculated. For the galaxy clusters and dwarf galaxies, we set their $\theta_{ROI}$ to the angular distances at which their  $D(\theta_{ROI})$ saturates. For the Andromeda galaxy, we limit the $\theta_{ROI}$ to $8^{\circ}$ in order to separate the ROI from the Galactic Plane.    

\renewcommand{\arraystretch}{1.1}
\begin{table}[t]
\centering
\resizebox{0.67\textwidth}{!}{
\begin{tabular}{|c|c|c|c|c|c|}
\hline
Source            & Type & $\alpha$[$^{\circ}$] & $\delta$ [$^{\circ}$] & $\theta_{ROI}$ [$^{\circ}$] & $\log_{10} (D_{ROI}/GeV/cm^{2})$ \\
\hline
\hline
Virgo             & \multirow{3}{*}{galaxy cluster}  & $186.63$  & $12.72$ & $6.11$ & $20.40$    \\
Coma              &                                  & $194.95$  & $27.94$ & $1.30$ & $19.17$    \\
Perseus           &                                  & $49.94$   & $41.51$ & $1.35$ & $19.15$    \\
\hline
Andromeda         & galaxy         & $10.68$   & $41.27$ & $8.00$ & $20.23$    \\
\hline
Draco             & \multirow{7}{*}{dwarf galaxy} & $260.05$ & $57.92$  & $1.30$ &   $18.97$ \\
Ursa Major II     &                               & $132.87$ & $63.13$  & $0.53$ &   $18.39$ \\ 
Ursa Minor        &                               & $227.28$ & $67.23$  & $1.32$ &   $18.13$  \\
Segue 1           &                               & $151.77$ & $16.08$  & $0.34$ &   $17.99$  \\
Coma Berenices    &                               & $186.74$ & $23.9$   & $0.34$ &   $17.96$ \\
Leo I             &                               & $152.12$ & $12.3$   & $0.45$ &   $17.92$  \\
Bo\"otes I        &                               & $210.03$ & $14.5$   & $0.53$ &   $17.90$  \\
\hline
\end{tabular} }
\caption{\textbf{Targets selected for the analysis.} The third and fourth columns from left show the right ascension ($\alpha$) and declination ($\delta$) of the sources, respectively, in equatorial coordinates for epoch J2000. The fifth column is the angular distance to the source center ($\theta_{ROI}$) corresponding to the ROI. The D-factor ($D_{ROI}$) is calculated up to $\theta_{ROI}$. The $\theta_{ROI}$ and $D_{ROI}$ values are determined using dark matter halo models presented in Refs.~\cite{SanchezConde:2011ap, Tamm:2012hw, Geringer-Sameth:2014}. In this analysis, the galaxy clusters and dwarf galaxies are stacked within the same source class.}
\label{table:sources}
\end{table}

\label{s:signal_and_background}

\section{Data Analysis} 
We perform a likelihood ratio test to search for an excess of signal neutrino events. 
The likelihood function is constructed as 
\begin{equation}
    L(n_{s}) = \prod^{N}_{i=1} \left[ \frac{n_{s}}{N}S(\alpha_{i}, \sin\delta_{i}, \sigma_{i}, E_{i}|n_{s}) + \left(1 - \frac{n_{s}}{N}\right) B(\alpha_{i}, \sin\delta_{i}, E_{i}) \right],
\label{eq:likelihood}
\end{equation}
where $S$ and $B$ are the signal PDF and background PDF, respectively. The product in Eq.~(\ref{eq:likelihood}) runs over $N$ events with index $i$. Each event is accounted by its right ascension ($\alpha_{i}$) and declination ($\delta_{i}$), known within an angular reconstruction error ($\sigma_{i}$), as well as energy ($E_{i}$). Given the information of $N$ events, the likelihood function is maximized with respect to the expected number of signal events ($n_{s}$). 

The test statistic is defined as
\begin{equation}
    TS = -2\ln\frac{L(n_{s} = 0)}{L(n_{s} = \hat{n}_{s})},
\end{equation}
where $\hat{n}_{s}$ is the best-fit value obtained by maximizing the likelihood function, $L(n_{s})$, under the alternative hypothesis ($n_{s}\geq0$). $L(n_{s}=0)$ is the likelihood function corresponding to the null hypothesis. Since we consider multiple dark matter signal cases corresponding to the different dark matter masses, decay channels, and the source groups, the global hypothesis test consists of multiple local hypothesis tests. A global p-value can be calculated from the local p-values by adopting the method presented in Ref.~\cite{IceCube:2010nca}. 

\begin{figure}[b]
    \centering
    \includegraphics[width=0.329\linewidth]{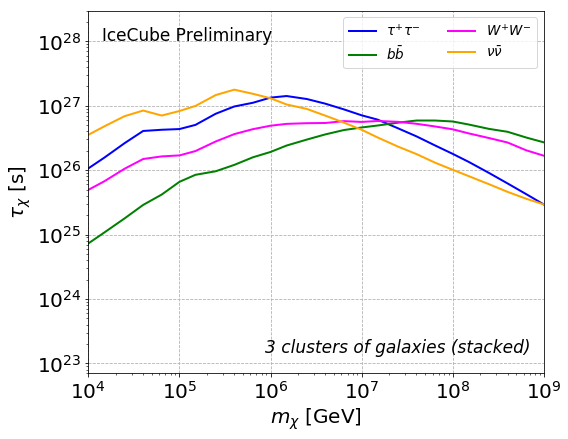}
    \includegraphics[width=0.329\linewidth]{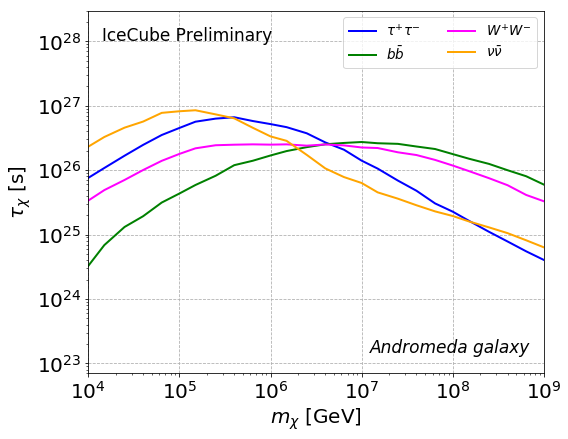}
    \includegraphics[width=0.329\linewidth]{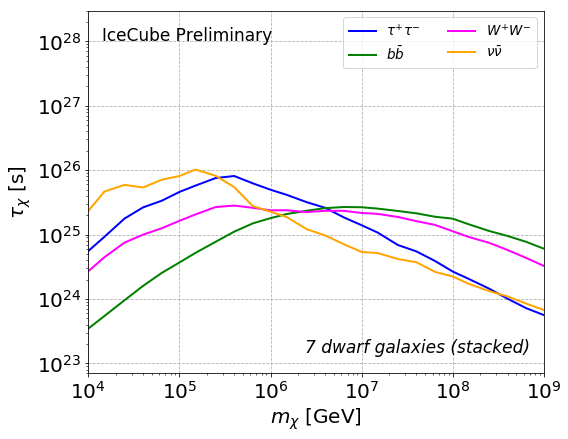}
    \caption{\textbf{90\% C.L. lower limits on the dark matter lifetime.} The three panels show the limits obtained using three different source groups: three galaxy clusters (left), the Andromeda galaxy (middle), seven dwarf galaxies (right). Properties of the sources are summarized in Table~\ref{table:sources}. In each panel the limits are shown for the four different decay channels considered in this analysis.}
\end{figure}

The signal and background PDFs are approximated as the product of a spatial part and an energy part 
such that $S = S_{S}(\alpha_{i}, \sin\delta_{i}, \sigma_{i})S_{E}(E_{i},\sin\delta_{i})$ and $B = B_{S}(\sin\delta_{i})B_{E}(E_{i},\sin\delta_{i})$. The energy part of the signal PDF is obtained by re-weighting neutrino Monte Carlo events by the expected signal neutrino spectrum and normalizing the resulting distribution. To obtain the spatial part, the angular distribution of the signal neutrino flux is calculated, with a grid of equal solid angle bins, and weighted by the declination-dependent signal acceptance of IceCube. The acceptance-corrected signal distribution is smeared for different angular reconstruction error bins ranging from $0^{\circ}$ to $3^{\circ}$ and then normalized. 
Since a signal PDF depends on the assumed dark matter mass, decay channel, and the source group, different signal PDFs are calculated for the different local hypothesis tests.

Backgrounds for this analysis include the atmospheric neutrino flux that is considered to be uniform in right ascension. The flux of astrophysical neutrinos originating from luminous matter in the Universe and extragalactic dark matter decay, in the foreground and background of the targets, is expected to be isotropic and thus uniform in right ascension. We exploit this property to estimate the distribution of background events at IceCube. We replace the right ascension of observed events in the data sample with random numbers from $0^{\circ}$ to $360^{\circ}$. Then the distribution of these scrambled events is assumed to represent the background event distribution. The neutrino flux from Galactic dark matter decay would not be uniform in right ascension. However, since the targets are located in the northern sky at higher declinations than $10^{\circ}$, the contribution from the Galactic dark matter decay to the scrambled data is considered to be negligible compared to the atmospheric and isotropic astrophysical neutrinos. The neutrino flux from luminous matter in the targets has not been detected and thus is difficult to model reliably. Hence, we assume that this background is negligible. As a consequence, if the observed significance is above 3$\sigma$, the result would not clearly indicate evidence for dark matter signals but would require further studies for identifying the source of the excess. Contributions from atmospheric muons are also considered negligible, as the analysis focuses on the northern sky.   

For this work, we use a well-established IceCube data sample that contains upward-going track-like events recorded from 2011 to 2022 for a total livetime of 11 years. Data samples based on essentially the same event selection were used in previous IceCube analyses~\cite{IceCube:2015qii, IceCube:2016umi}. For details regarding the IceCube detector see Ref.~\cite{IceCube:detector}. The event selection achieves sub-degree angular resolution for energies above a few TeV, which is preferred in searches for neutrinos from point-like or small extended sources. The events in this sample cover declination from $-5^{\circ}$ to $90^{\circ}$, where $90^{\circ}$ coincides with the nadir at IceCube's location. The sample is almost free of atmospheric muon backgrounds, as the muons from the northern sky are absorbed by the Earth. For declinations from $-5^{\circ}$ to $0^{\circ}$ the overburden of the Antarctic ice sufficiently attenuates the atmospheric muon flux.

\begin{figure}[t]
    \centering
    \includegraphics[width=0.7\linewidth]{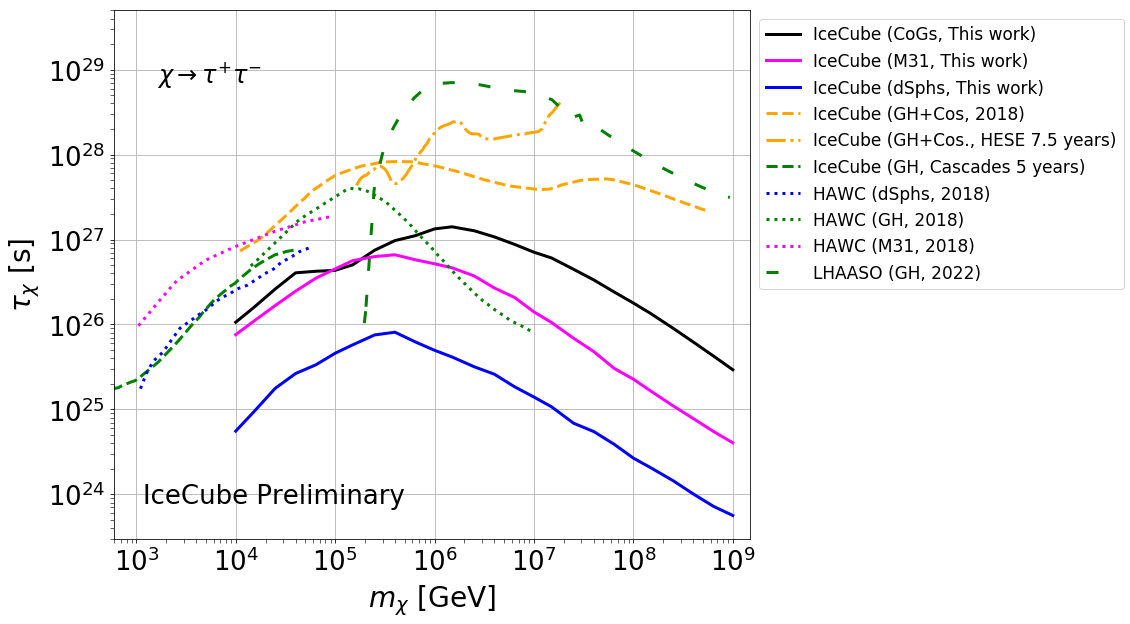}
    \caption{\textbf{Comparison of lower limits on the dark matter lifetime for the $\tau^{+}\tau^{-}$ channel.} The solid lines represent the limits calculated in this work using the three different source groups. The other lines are the limits from recent dark matter searches with IceCube~\cite{IceCube:EPJC2018DM, IceCube:HESE7.5DM, IceCube:NuLine2023ies}, HAWC~\cite{HAWC:GH_2018, HAWC:dSphs, HAWC:M31}, and LHAASO~\cite{LHAASO:2022yxw}. The line colors indicate different targets used for the analyses: clusters of galaxies (black), dwarf spheroidal galaxies (blue), M31  (magenta), the Galactic Halo (green), and a combination of the Galactic Halo and cosmological dark matter (orange). The confidence levels associated with the limits are 90\% for the IceCube and LHAASO results, and 95\% for the HAWC results.}
\end{figure}

\begin{figure}[t]
    \centering

    \includegraphics[width=0.69\linewidth]{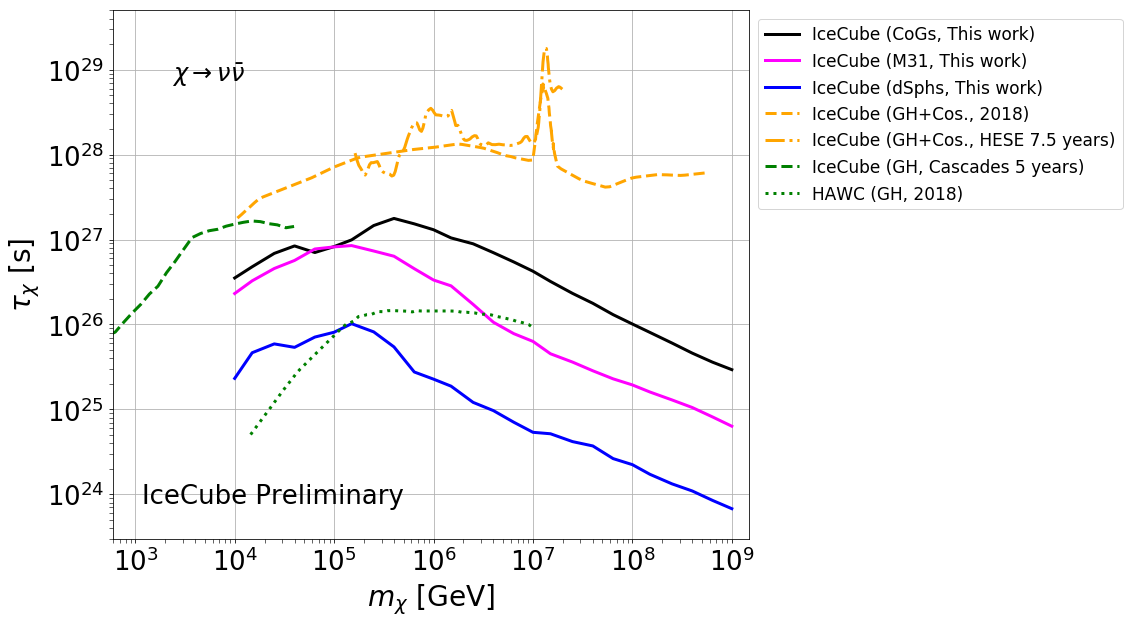}
    \caption{\textbf{Comparison of lower limits on the dark matter lifetime for the $\nu\bar{\nu}$ channel.} The solid lines represent the limits calculated in this work using the three different source groups. The other lines are the limits from recent dark matter searches with IceCube~\cite{IceCube:HESE7.5DM, IceCube:NuLine2023ies, IceCube:EPJC2018DM} and HAWC~\cite{HAWC:GH_2018}. The line colors indicate different targets used for the analyses: clusters of galaxies (black), dwarf spheroidal galaxies (orange), M31 (magenta), the Galactic Halo (green), and a combination of the Galactic Halo and cosmological dark matter (orange). The confidence levels associated with the limits are 90\% for the IceCube results and 95\% for the HAWC results.}
\end{figure}

\section{Results}
By testing the multiple local hypotheses, we obtain the most significant local p-value of 0.013 which corresponds to a significance of $2.2\sigma$. Due to the trials factor associated with repeating multiple local hypotheses tests, the global significance should be smaller than $2.2\sigma$. Hence, we conclude that no evidence from dark matter decay in the targets is found, and derive lower limits on the dark matter lifetime at 90\% confidence level. These limits are presented in Figure 2. The best limits from this work are obtained by stacking the three galaxy clusters, and the next best limits by using the Andromeda galaxy. This is as expected from their $D_{ROI}$ values presented in Table~\ref{table:sources}. The limits for the $\tau^{+}\tau^{-}$ and $\nu\bar{\nu}$ channels are compared to the results from recent dark searches in Figure 3 and 4, respectively. In both figures, it can be seen that the recent dark matter searches provide more stringent limits than this work. The recent IceCube analyses looked for signals from dark matter decay in the Galactic Halo ("GH") or both the Galactic and cosmological dark matter decay ("GH + Cos."), and $D_{ROI}$ is substantially larger for the Galactic Halo than those of the selected targets for this analysis. The high signal strength from the Galactic Halo would also explain the competitive limits from Galactic Halo analyses with LHAASO and HAWC shown in Figure 3. The HAWC analyses of the Andromeda galaxy ("M31") and dwarf spheroidal galaxies ("dSphs") provide stronger limits on the dark matter lifetime than the corresponding limits from this work as shown in the same figure. For the energy range covered by HAWC, the gamma-ray telescope has a superior effective area and angular resolution than IceCube. While the large field of view of IceCube is advantageous to look for signals from the Galactic Halo, for limited ROIs HAWC would be more sensitive. However, the presented analysis can complement the recent IceCube analyses by using different types of targets, and gamma-ray experiments by using neutrinos.

\section{Conclusions and Outlook}\label{s:conclusions}

The presented analysis is the first search for neutrinos from dark matter decay in galaxy clusters and galaxies. We found no evidence for dark matter decay in the selected targets and derived lower limits on the dark matter lifetime. Limits from recent decaying dark matter searches with neutrinos and gamma-ray experiments are significantly better than the corresponding limits from this work. However, this analysis can complement those neutrino and gamma-ray analyses by using either different types of targets or different messenger particles, or both. 

Given that no dark matter signal is found from the presented analysis, limits on the dark matter lifetime can also be derived for the individual galaxy clusters and dwarf galaxies, without increasing the trials factor. Limits can also be calculated for decay channels that are highly correlated with those considered for the main analysis. For examples, the expected neutrino spectra from the $Z^{0}Z^{0}$ channel are similar to those from the $W^{+}W^{-}$ channel, and the $\nu_{e}\bar{\nu}_{e}$, $\nu_{\mu}\bar{\nu}_{\mu}$, and $\nu_{\tau}\bar{\nu}_{\tau}$ channels are highly correlated with the $\nu\bar{\nu}$ channel. These limits will be further discussed in a future publication.

Future IceCube analyses of this kind would achieve improved sensitivities to the dark matter lifetime. The IceCube-Gen2 is a proposed high-energy extension to the current IceCube detector that is expected to have a factor of eight larger effective volume than IceCube. The increased volume would not only increase the event rate but also improve the angular resolution for track-like events~\cite{IceCube-Gen2:2023icrc_overview}. 

\bibliographystyle{ICRC}
\bibliography{references}

%

\clearpage

\section*{Full Author List: IceCube Collaboration}

\scriptsize
\noindent
R. Abbasi$^{17}$,
M. Ackermann$^{63}$,
J. Adams$^{18}$,
S. K. Agarwalla$^{40,\: 64}$,
J. A. Aguilar$^{12}$,
M. Ahlers$^{22}$,
J.M. Alameddine$^{23}$,
N. M. Amin$^{44}$,
K. Andeen$^{42}$,
G. Anton$^{26}$
% ...
\\
\\
% ...
$^{12}$ Universit{\'e} Libre de Bruxelles, Science Faculty CP230, B-1050 Brussels, Belgium \\
% ...
$^{17}$ Department of Physics, Loyola University Chicago, Chicago, IL 60660, USA \\
$^{18}$ Dept. of Physics and Astronomy, University of Canterbury, Private Bag 4800, Christchurch, New Zealand \\
% ...
$^{22}$ Niels Bohr Institute, University of Copenhagen, DK-2100 Copenhagen, Denmark \\
$^{23}$ Dept. of Physics, TU Dortmund University, D-44221 Dortmund, Germany \\
% ...
$^{26}$ Erlangen Centre for Astroparticle Physics, Friedrich-Alexander-Universit{\"a}t Erlangen-N{\"u}rnberg, D-91058 Erlangen, Germany \\
% ...
$^{40}$ Dept. of Physics and Wisconsin IceCube Particle Astrophysics Center, University of Wisconsin{\textendash}Madison, Madison, WI 53706, USA \\
% ...
$^{42}$ Department of Physics, Marquette University, Milwaukee, WI, 53201, USA \\
% ...
$^{44}$ Bartol Research Institute and Dept. of Physics and Astronomy, University of Delaware, Newark, DE 19716, USA \\
% ...
$^{63}$ Deutsches Elektronen-Synchrotron DESY, Platanenallee 6, 15738 Zeuthen, Germany  \\
$^{64}$ Institute of Physics, Sachivalaya Marg, Sainik School Post, Bhubaneswar 751005, India \\
% ...

\subsection*{Acknowledgements}

\noindent
The authors gratefully acknowledge the support from the following agencies and institutions:
USA {\textendash} U.S. National Science Foundation-Office of Polar Programs,
U.S. National Science Foundation-Physics Division,
U.S. National Science Foundation-EPSCoR,
Wisconsin Alumni Research Foundation,
Center for High Throughput Computing (CHTC) at the University of Wisconsin{\textendash}Madison,
Open Science Grid (OSG),
Advanced Cyberinfrastructure Coordination Ecosystem: Services {\&} Support (ACCESS),
Frontera computing project at the Texas Advanced Computing Center,
U.S. Department of Energy-National Energy Research Scientific Computing Center,
Particle astrophysics research computing center at the University of Maryland,
Institute for Cyber-Enabled Research at Michigan State University,
and Astroparticle physics computational facility at Marquette University;
Belgium {\textendash} Funds for Scientific Research (FRS-FNRS and FWO),
FWO Odysseus and Big Science programmes,
and Belgian Federal Science Policy Office (Belspo);
Germany {\textendash} Bundesministerium f{\"u}r Bildung und Forschung (BMBF),
Deutsche Forschungsgemeinschaft (DFG),
Helmholtz Alliance for Astroparticle Physics (HAP),
Initiative and Networking Fund of the Helmholtz Association,
Deutsches Elektronen Synchrotron (DESY),
and High Performance Computing cluster of the RWTH Aachen;
Sweden {\textendash} Swedish Research Council,
Swedish Polar Research Secretariat,
Swedish National Infrastructure for Computing (SNIC),
and Knut and Alice Wallenberg Foundation;
European Union {\textendash} EGI Advanced Computing for research;
Australia {\textendash} Australian Research Council;
Canada {\textendash} Natural Sciences and Engineering Research Council of Canada,
Calcul Qu{\'e}bec, Compute Ontario, Canada Foundation for Innovation, WestGrid, and Compute Canada;
Denmark {\textendash} Villum Fonden, Carlsberg Foundation, and European Commission;
New Zealand {\textendash} Marsden Fund;
Japan {\textendash} Japan Society for Promotion of Science (JSPS)
and Institute for Global Prominent Research (IGPR) of Chiba University;
Korea {\textendash} National Research Foundation of Korea (NRF);
Switzerland {\textendash} Swiss National Science Foundation (SNSF);
United Kingdom {\textendash} Department of Physics, University of Oxford.

\end{document}